# Phenomenology of disruptive breakup mechanism of a levitated evaporating emulsion droplet

D. Chaitanya Kumar Rao[*] and Saptarshi Basu[§]

*Department of Mechanical Engineering, Indian Institute of Science, Bangalore – 560012, India*

**Abstract**

Atomization of emulsion droplets is ubiquitous across a variety of application domains ranging from NextGen combustors to fabrication of biomedical implants. An understanding of the atomization mechanism in emulsions can result in a paradigm shift in customized designs of efficient systems, be it in energy or biotechnology sectors. In this paper, we specifically study the breakup mechanism of an evaporating contact-free (acoustic levitation) emulsion droplet (water-oil) under external heating. Three distinct regimes are observed during the lifespan of the evaporating droplet. Initially, the droplet diameter regresses linearly with time, followed by vapor bubble nucleation due to a significant difference in the boiling temperature among the components of the emulsion. The collapse of this bubble results in a high-intensity breakup of the droplet leading to the propulsion of residual liquid in the form of a crown-like sheet. The area of the expanding crown varies linearly with the square of the time. It is hypothesized that the expansion of the liquid sheet centrifuges the larger water sub-droplets towards the edge, resulting in unique spatial segregation. Subsequently, we report the first observation of complex patches (representing water sub-droplets) and the rupture of the thin sheet adjacent to patches into holes (with hole growth rate ranging from 1.2 to 1.4 m/s) in the context of an evaporating isolated emulsion droplet. The hole formation results in the creation of ligaments which undergo breakup into secondary droplets with Sauter mean diameter (SMD) ~ 50 µm.

# 1    Introduction

Droplet deformation and atomization are of considerable significance to a wide gamut of applications like fuel combustors, ink-jet printing, and medical technology [1–3]. In particular, droplet atomization characteristics in combustors play a significant role in governing the combustion efficiency and pollutant emissions [4]. Similarly, the performance of an ink-jet printer depends on the size distribution of droplets [1]. The atomization phenomenon is a complicated interfacial process governed by parameters such as surface tension, viscosity, external forces (inertial, acoustic, and electromagnetic force), and non-dimensional groups thereof. Depending on the non-dimensional parameters such as Weber number, various types of breakup have been reported in the literature viz., bag, oscillatory, and catastrophic breakup, to name a few [1].

[*]chaithanyadevv@gmail.com

[§]sbasu@iisc.ac.in

Emulsion or colloidal droplets pose a unique set of challenges in terms of droplet deformation and breakup. The breakup phenomenon has been extensively reported in multi-component and emulsion droplets across various experimental configurations [5–22] and numerical investigations [23–27]. Significant experimental work on the emulsion droplet breakup is performed on the combustion of pendant droplets [5,10,15,18], free-falling droplets [6,8,28]. The most common techniques to investigate heated emulsion droplets is through intrusive techniques (by suspending the droplet on thin fiber or thermocouple) [5,11] and low or non-intrusive methods (such as freely falling approach under microgravity conditions) [6,8,15]. In the case of the intrusive method, the experimentation is affected by the interaction of a thin fiber or thermocouple with the droplet. This interaction can result in the formation of unwanted heterogeneous nucleation sites inside the droplet [5]. In contrast, the non-intrusive free-falling method helps in the measurement of parameters like burning rate constant, sooting tendency, and flame diameter. However, due to the intricacies associated with the free-falling method, it is rather challenging to obtain the qualitative data about processes such as the small-scale instabilities during the evaporation and combustion process. On the other hand, the relatively simplistic and contact-free nature of acoustic levitation [29,30], along with its uncomplicated experimental apparatus, allows one to capture short spatio-temporal instabilities such as Rayleigh-Taylor instability with reduced difficulty [29]. In-depth studies on the atomization of emulsions are rare, especially under contact-free conditions like acoustic levitation.

On the other hand, a limited number of studies, particularly in the domain of drop impacts have been carried out on the breakup mechanisms of emulsion droplets [31,32]. Vernay et al. [31] investigated the breakup of oil-in-water emulsion droplet impacting on a solid surface. It was shown that the radially expanding liquid sheet formed due to the impingement is characterized by a network of holes bounded by unstable liquid filaments due to the formation of dark patches. The occurrence of this breakup mechanism has also been reported for droplets impacting on a pool of a liquid with lower surface tension, characterized by a crown-like sheet [33,34]. It was proposed that the crown breakup results from a spray of fine droplets ejected from the thin low-viscosity film on the solid. These droplets interact with the crown forming dark spots with low surface tension, which subsequently leads to the patch thinning and further hole formation [33]. In addition, the growth rate of the patch thinning as well as hole formation was quantified. This crown sheet is similar to the ones observed in the context of impact of droplets onto thin films [35–37], dry solid surfaces [38], and laser pulse impact [39,40]. Recently, the formation of the crown-like sheet has also been reported in burning multi-component pendant droplets [41] and laser-induced cavitation of levitated droplets [29].

Although a few studies have been focused on understanding the breakup mechanisms through patch and hole formation, there has been no such investigation on evaporating acoustically levitated emulsion droplets. The implications of such a study are significant as the individual droplets represent the sub-grid level component of any evaporating or burning fuel spray, and their breakup mimics the essential atomization physics in a typical combustor.

In this work, we describe the breakup mechanism and associated physics for an isolated contact-free emulsion (water-in-n-decane) droplet under external heating. Primarily, we delineate three distinct regimes during the lifespan of the evaporating emulsion droplet. The overview of the droplet evaporation and breakup phenomenon is presented in section 3.1. We show that during the breakup process, the area of the expanding crown varies linearly with the square of the time (section 3.2). We also present the hypothesis of the migration of water sub-droplets due to centrifuging effect of the expanding crown, resulting in unique spatial segregation of sub-droplets from a randomly well-stirred dispersion (section 3.3). Finally, we describe the mechanism of patch and hole formation in the liquid sheet, which is first of its kind in the context of a single isolated contact-free droplet (section 3.3). Despite the stochasticity and complexity involved in capturing the complex patches and holes, a probable hypothesis of its occurrence is proposed in this study.

## 2 Experimental methodology

In the present experiments, the initial diameter of the emulsion droplet is maintained at $450 \pm 50$ µm. The emulsion consists of 20% (v/v) water while 2.5% (v/v) surfactant (SPAN 80) is used to stabilize the emulsion. The composition of emulsion used in the present work is selected following the experimental investigation on the combustion of water in decane droplets by Kim and Baek [42]. Similar to their work, the breakup of droplets via homogeneous nucleation occurs in 10%, 20%, and 30% volume water concentrations. However, the breakup of droplets through the formation of patches and holes occurs only in droplets with 20% water concentration. In the other mixtures, however, the breakup mechanisms are qualitatively as well as quantitatively different, and therefore, they are not in the scope of present work (see Ref. [43]). The properties of the tested liquids are listed in Table 1. The mixture is sonicated for 15 minutes with continuous stirring. The experiments are conducted within 30 minutes from the sonication of the mixture. The emulsion droplets are levitated at the pressure nodes of the acoustic levitator (Tec5, 100 kHz, 154 dB) and externally heated using a tunable $CO_2$ laser (Synrad 48, beam diameter of 3.5 mm, and wavelength of 10.6 μm) at an irradiation intensity of 0.99 MW/m$^2$. The effective laser flux is lower than the actual laser power because of factors such as (a) absorption coefficient, (b) divergence of laser beam, and (c) scattering loss. However, except for absorption coefficient, the other factors generally depend on radiation wavelength and geometry of the droplet [44]. With the consideration that the levitated droplets are nearly spherical in shape before the occurrence of breakup (as will be seen in the results), it can be assumed that these effects are uniform for all liquids when heated with the same monochromatic source. The droplet breakup process is captured using a high-speed camera (Phantom Miro Lab110) coupled with a 5X zoom lens. The high-speed images are obtained at a spatial resolution of 0.16 μm/pixel and a frame rate of 20,000 fps while the exposure is maintained at 10 μs. The non-intrusive droplet surface temperature is measured using an infrared camera (FLIR SC5200: pre-calibrated for a standard emissivity of 1) at 500 fps with a spatial resolution of 12.4 μm/pixel. The emissivity for water is in the range of 0.95–0.98 [45,46], and the error in temperature for this range of emissivity is 0.03 °C, which is assumed to be insignificant. In the present work, the emissivity is kept at 0.95. The captured IR images are processed by ALTAIR software in order to obtain the droplet temperature profile. First, the greyscale images were converted to rainbow palette available in the software. Subsequently, the temperature data was obtained by defining a linear region of interest along the diameter of the droplet in each IR image, and the average temperature on the droplet surface was measured. The core temperature of the drop is expected to be larger than the surface due to the penetration of radiation inside the droplet [47]. Due to the relatively small depth of field, the IR images were sometimes out of focus. Therefore, only the images which were in focus were considered for further analysis. In the present experiments, we did not have a straightforward method of determining the transmissivity and the reflectivity of the liquid. When the emulsion droplet and pure decane were levitated, both of them showed the same temperature (using emissivity value of 0.95, without heating) as ambient air, which was measured through a thermocouple. Therefore, together with the effects of potential transmissivity and reflectivity changes, the maximum error in measurements of surface temperature may be approximated to be ± 1 °C. A similar observation was reported for nano-particle laden droplets using an emissivity value of 0.98 [44]. Moreover, considering the levitated droplets are nearly spherical and are of constant volume before the occurrence of breakup (which will be shown in the experimental results), it can be assumed that these effects are uniform for all liquids when heated with the same monochromatic source.

**Table 1. Properties of tested liquids.**

| Property | n-Decane | Water |
|---|---|---|
| Boiling point (K) | 447 | 373 |
| Density at 105 °C (Kg/m$^3$) | 664[+] | 944* |
| Surface tension at 105 °C (mN/m) | 16[$] | 58.9* |
| Viscosity at 105 °C (cP) | 0.34[$] | 0.26* |
| Heat of vaporization (kJ/kg) | 276[§] | 2260[§] |

[+]Density of decane is obtained from [48]

*Saturated liquid properties [49]

[$]Surface tension and viscosity of decane are obtained from [49]

[§]Heat of vaporization of liquids are obtained from [42]

The high-speed and IR cameras are synchronized with the laser using a delay generator. Readers can refer to Ref. Raju et al. [50] for additional details of the experimental apparatus. The diameter of the water sub-droplets in the emulsion is measured using an optical microscope (Olympus BX-51, Japan) before conducting the experiments. Some form of coalescence and phase separation may occur during the heating process. However, due to the opaqueness of the droplet (shown in the results section), it is rather difficult to discern any convection and coalescence or phase separation inside the droplet during the evaporation process. The regression of the droplet diameter is obtained using an in-house MATLAB code [41]. The velocity, acceleration, and rotation of the deforming parent droplet, as well as the patch and hole diameter, are measured using an image analysis platform (Image-Pro Plus). It is important to note that the position of the bubble at the onset of breakup and the direction of vapor thrust can be stochastic. This stochastic nature of nucleation and bubble bursting can be attributed to the multi-component and multiphase flow involved during the evaporation of emulsions [22]. Moreover, the levitated droplet continuously oscillates due to the presence of the acoustic field, which makes it almost impossible for the bubble to burst and eject the vapor from a single location for all the test cases. The quantitative data in the present work were obtained such that the patches and holes are clearly visible, in which vapor thrust direction varies from 0 to 55 degrees with respect to the horizontal axis of the image. Because the observation of patches is extremely sensitive to optical conditions, only four test cases were considered to obtain the quantitative data (such as centrifugal force, crown area, and patches and hole diameter). However, for ligament and secondary droplet size distribution, a total of 8 cases were considered. Although the patches and holes were observed in all the test cases, it is important to note that the patches and holes are not always in focus during the sheet expansion. Therefore, only the cases where the patches are in focus are considered and the patches which are not entirely perpendicular to the camera are neglected. For the determination of patches, holes, ligament, and secondary droplet diameter, the experimental and calibration uncertainties are found to be no more than 5%. The area of the expanding crown is obtained through two-dimensional projections of the crown considering axisymmetric expansion, and thus the data do not resolve the curvature of the crown. The maximum error in estimating the crown area and centrifugal force is of the order of 15%. The experiments are repeated at least ten times for ensuring repeatability and the patches and holes were observed in all the test cases.

## 3  Results and discussion

### 3.1  Overview of droplet evaporation and breakup

For a pure decane case, the droplet undergoes smooth regression without any disruption, as evident in Fig. 1(a). As seen in Fig. 1(a), after the preheating regime, the droplet diameter regresses linearly with time until it is completely evaporated. On the other hand, the droplet regression data acquired from high-speed imaging and temperature profiles obtained from infrared thermography indicate that the heating process of water/decane emulsion droplet consists of three distinct regimes for all the droplets studied, namely, evaporation (Regime I), nucleation and bubble collapse (Regime II), and sheet fragmentation through patch and hole formation (Regime III). Similar regimes (preheating and steady evaporation, bubble formation, and breakup) have been extensively reported in the literature during the combustion of multi-component miscible droplets [13,14], emulsion droplets [42] and nanoparticle laden droplets [51]. However, the pattern of breakup in the present work is entirely different from the investigations mentioned above, owing to distinct experimental conditions.

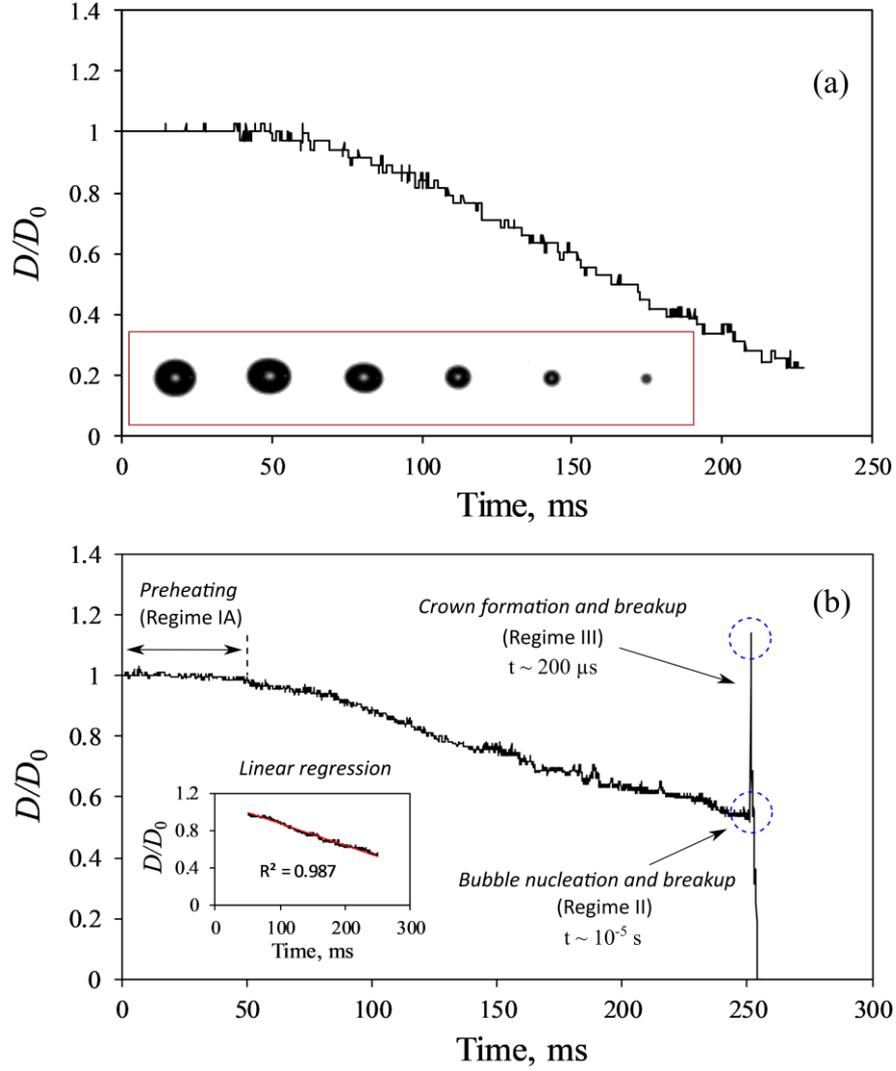

**Fig. 1. (a) Variation of normalized diameter ($D/D_0$) with time for a typical pure decane droplet. Inset shows the smooth evaporation and diameter regression of the droplet with frames separated by 50 ms interval. (b) Variation of normalized diameter ($D/D_0$) with time for a water/decane emulsion droplet representing different regimes during the evaporation and subsequent breakup of a typical droplet. The inset shows the linear regression of normalized droplet diameter with time.**

The presence of surfactant does not play any role in the fragmentation process as observed in our preliminary experiments with oil/surfactant droplets, which is possibly due to a significantly lower proportion of surfactant in the emulsion (2.5% v/v) and miscibility of oil/surfactant mixtures. Moreover, the conclusion of the negligible influence of the surfactant was drawn based on the investigation by Kim and Baek [42] on the combustion of water-in-decane emulsion droplets on a thermocouple. For a similar proportion of surfactant (2% volume), it was shown that the surfactant evaporation/combustion occurs only after the breakup phenomenon. Moreover, the temperature required for the evaporation of the surfactant is significantly high (>350 °C) [42], which is not achieved in the present work. Nevertheless, further work is needed to discuss the influence of surfactant on breakup in more detail.

Figure 1b shows the temporal evolution of droplet diameter for a typical emulsion. Regime I is further characterized into two stages, i.e., initial droplet preheating ($0 \leq t \lesssim 50$ ms) and droplet diameter regression ($50 \lesssim t \lesssim 250$ ms). During the preheating regime (Regime IA), the droplet diameter remains nearly constant which is followed by the linear regression of droplet diameter ($D$) with time ($t$) (Regime IB). This linear relation signifies that the droplet evaporation in the present case is limited by the magnitude of external heating [50].

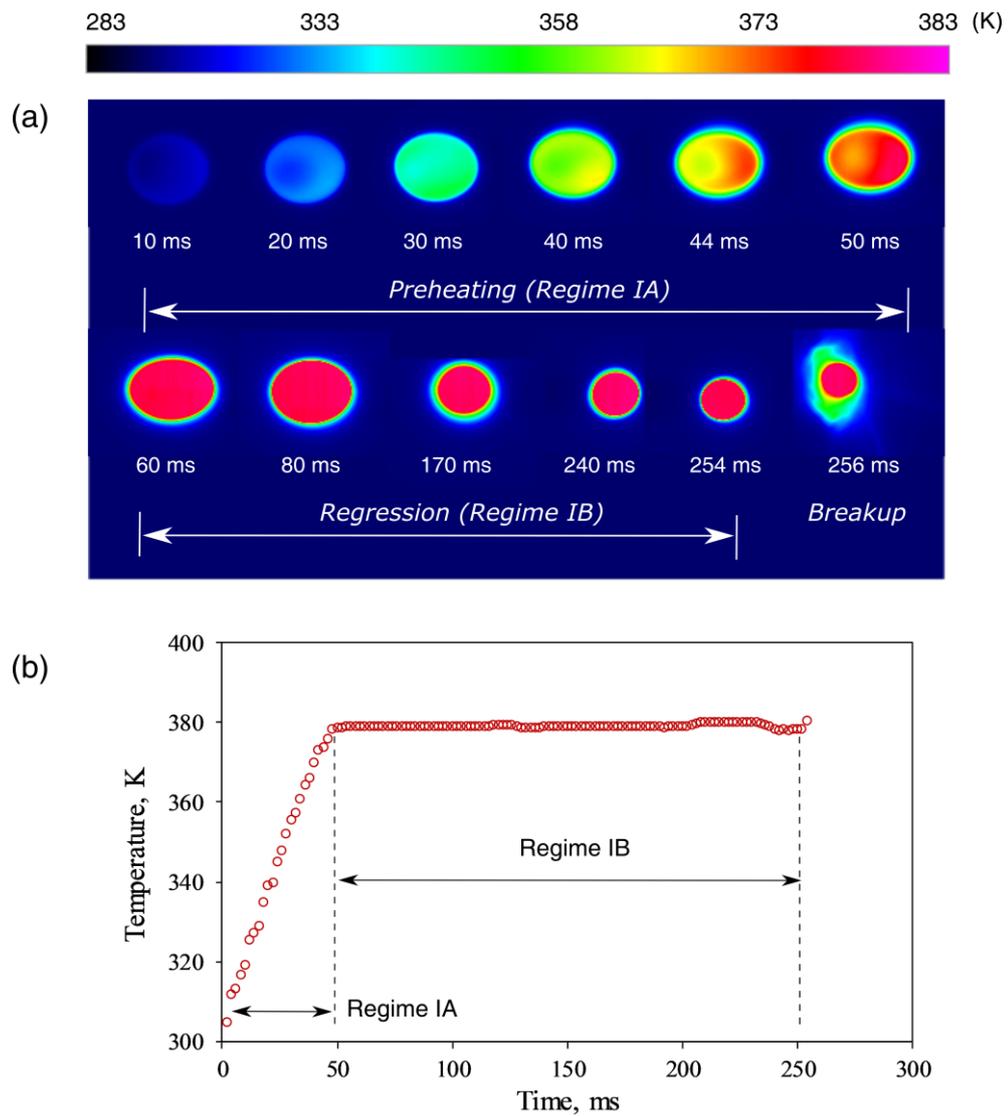

Fig. 2. (a) Infrared images showing the droplet surface temperature distribution during the preheating and regression regimes. (b) Temporal evolution of the droplet surface temperature. The droplet is heated at an irradiation intensity of 0.99 MW/m$^2$.

The preheating and the linear regression can also be delineated from the maximum droplet temperature profile (Fig. 2(a) and (b)). The droplet temperature profile shows an increase in surface temperature for the initial 50 ms of the heating process and consequently attains a nearly constant value (105 °C) until its breakup at approximately 250 ms. Therefore, further analysis of droplet breakup dynamics has been done using the fluid properties at 105 °C (see Table 1). During the final stage of Regime I, the water sub-droplets present in the parent droplet are superheated due to significant volatility difference between the liquid components since the boiling temperature ($T_b$) of water ($T_b$ = 373 K) is significantly lower than decane ($T_b$ = 447 K). This superheating results in the formation of a high-pressure water vapor bubble inside the parent droplet (Regime II), as illustrated in Fig. 3(a). The timescale of Regime II is significantly shorter (~ $10^{-5}$ s) than other regimes since it is associated with the nucleation and rapid breakup of the vapor bubble.

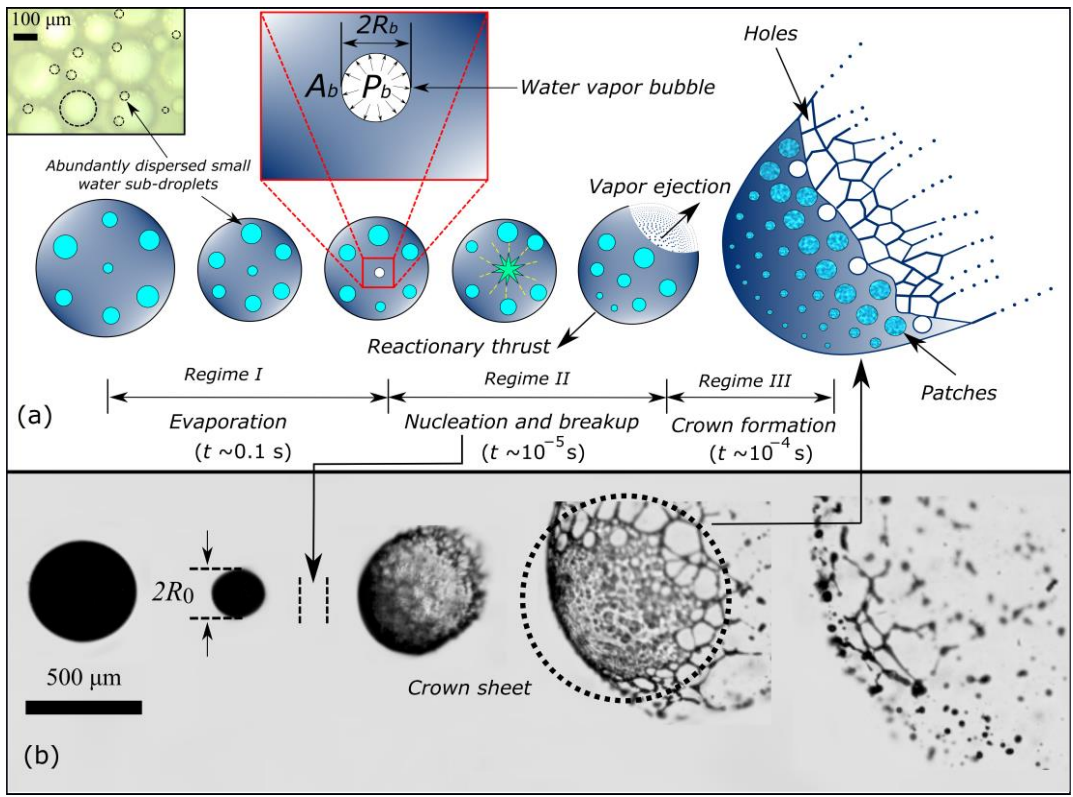

**Fig. 3. (a) Schematic of the evolution of water-in-decane emulsion droplet under external heating. Regime I corresponds to initial evaporation of the droplet, Regime II represents nucleation and the subsequent breakup of the vapor bubble, and Regime III signifies crown-like sheet formation and development of patches and holes. The inset shows the observation of components of the emulsion under an optical microscope. (b) Experimental observation of patches and holes during the crown breakup of water-in-decane emulsion droplet. The first frame represents initial droplet and the second frame corresponds to the droplet at the pre-breakup instant. The last four frames are separated by 50 μs.**

The ratio of droplet diameter at the onset of breakup to the initial diameter ranges from 0.45 − 0.55. It can be contemplated that only a single bubble breakup event leads to droplet fragmentation due to significantly short nucleation and breakup timescales. Moreover, the formation of a nearly axisymmetric sheet in the post-breakup phase signifies the occurrence of a single breakup event. The onset of breakup was identified by observing the first appearance of droplet deformation, which occurs due to the breakup of the bubble, considering the negligible time between the actual bubble breakup and the observation of droplet deformation. The bursting of this single vapor bubble leads to the disintegration of the parent droplet. Initially, a cloud of daughter droplets, along with the water vapor, is ejected from the parent droplet. Subsequently, the residual liquid propels in the direction opposite to the ejected daughter droplets in the form of a crown-like sheet. The first frame in Fig. 3(b) corresponds to initial droplet diameter ($D_0$) while the remaining frames (from left to right, at 50 µs interval) correspond to the droplet at the pre-breakup instant, initiation of fragmentation, development of patches and holes in the crown, and complete sheet fragmentation.

### 3.2 Crown sheet expansion

During the propulsion of the residual liquid sheet, the laser radiation pressure and thermal-radiation pressure caused by droplet surface heating are negligible [39,52]. Therefore, the sheet motion is primarily a result of the recoil force due to the collapse of the vapor bubble. The resulting vapor recoil pressure deforms the droplet propelling it backward at high acceleration (~ $6 \pm 2 \times 10^4$ m/s$^2$). This deformation occurs at the inertial time scale [39], $\tau_i = R_0/U \sim 10^{-5}$ s, where $R_0$ is the radius of the droplet at the pre-breakup instant (110 µm), and $U$ is the maximum velocity of the sheet (~5 – 8 m/s).

Regime III describes the motion of the crown accompanied by its expansion in the radial direction. The radial expansion of the crown sheet (with $R_0$ = 110 µm) is directly associated with liquid surface tension, which attempts to inhibit deformation at capillary timescale [39], $\tau_c = \sqrt{\rho R_0^3/\sigma} = 235$ µs, where $\rho$ and $\sigma$ are the density and surface tension of the liquid respectively. The order of the capillary time agrees with the experimental sheet breakup timescale (~ 150 µs) (see Fig. 3(b)). Here, the relative importance of viscosity over the surface tension is negligible since the Ohnesorge number, $Oh < 0.05$, where $Oh = \mu/\sqrt{\rho h \sigma}$. Here $\mu$ is the viscosity of the liquid and $h$ is the thickness of the liquid sheet. The radial expansion of the sheet is described using normalized area ($A^*$) as a function of normalized time ($t/\tau_c$). The normalized crown area is defined as $A^* = A_{crown}/A_{drop}$, where $A_{crown}$ is the crown area, and $A_{drop}$ is the droplet area at the pre-breakup instant. The normalized area is observed to vary as the square of normalized time (i.e., $A^* \propto (t/\tau_c)^2$), as shown in Fig. 4(a). Additionally, the sheet decelerates in the present case at a constant rate during expansion (see inset of Fig. 4(a)). These observations of sheet expansion are consistent with the results reported in the framework of sheet expansion due to laser-pulse impact [39,40].

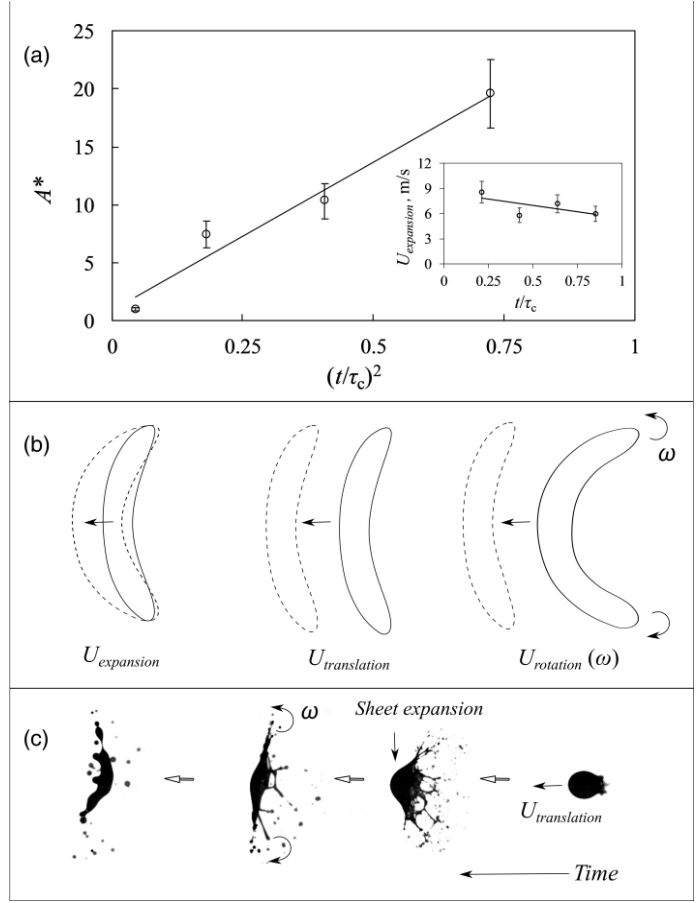

**Fig. 4. (a) Variation of the normalized area ($A^* = A_{crown}/A_{drop}$) with normalized time ($t/\tau_c$) during the crown breakup. The area of the expanding crown is obtained through two-dimensional projections of the crown considering axisymmetric expansion, and thus the data do not resolve the curvature of the crown. The total error in estimating the crown area due to the uncertainty in pixels is of the order of 15%. The normalized area varies as the square of normalized time (i.e., $A^* \propto (t/\tau_c)^2$). The right inset shows the variation of expansion velocity ($U_{expansion}$) of the crown sheet with the normalized time ($t/\tau_c$). (b) Schematic showing three fundamental motions that constitute the evolution of the crown sheet. (c) Experimental observation of the evolution of the crown sheet depicting the translational motion, expansion, and rotation of the sheet.**

In the present case, immediately after bubble bursting, the total kinetic energy ($K.E._{total}$) of the expanding sheet may be approximated as,

$$K.E._{total} = \underbrace{\frac{1}{2} m_{crown} U^2_{expansion}}_{K.E._{expansion}} + \underbrace{\frac{1}{2} m_{crown} U^2_{translation}}_{K.E._{translational}} + \underbrace{\frac{1}{2} I_{crown} \omega^2}_{K.E._{rotational}}. \tag{1}$$

Here, $m_{crown}$ is mass of the crown sheet, $I_{crown}$ is the mass moment of inertia and $\omega$ is the angular velocity of the sheet. $U_{expansion}$ and $U_{translation}$ are the expansion and translational velocity of the sheet respectively. It can be contemplated that the influence of $K.E._{translational}$ and $K.E._{rotational}$ on the sheet expansion is minimal compared to $K.E._{expansion}$ (see Fig. 4(b)). For a droplet with $R_0 = 110$ µm, the order of magnitude of both $K.E._{translational}$ and $K.E._{expansion}$ ranges from $\sim 10^{-8}$ to $10^{-7}$ J, while $K.E._{rotational}$ is $\sim 10^{-11}$ J. However, as will be discussed in section III C, the rotational velocity ($\omega$) seems to play a significant role in the migration of water sub-droplets and subsequent hole formation. Figure 4(c) depicts the side view of the evolution of crown sheet, where the translational motion, expansion, and rotation of the sheet are clearly demarcated.

## 3.3 Patches and holes

The formation and growth of dark patches occur on the sheet during its expansion and thinning due to the surface tension gradient [53]. Similar patches have been reported to form due to the spreading of oil sub-droplets on water [31,32].

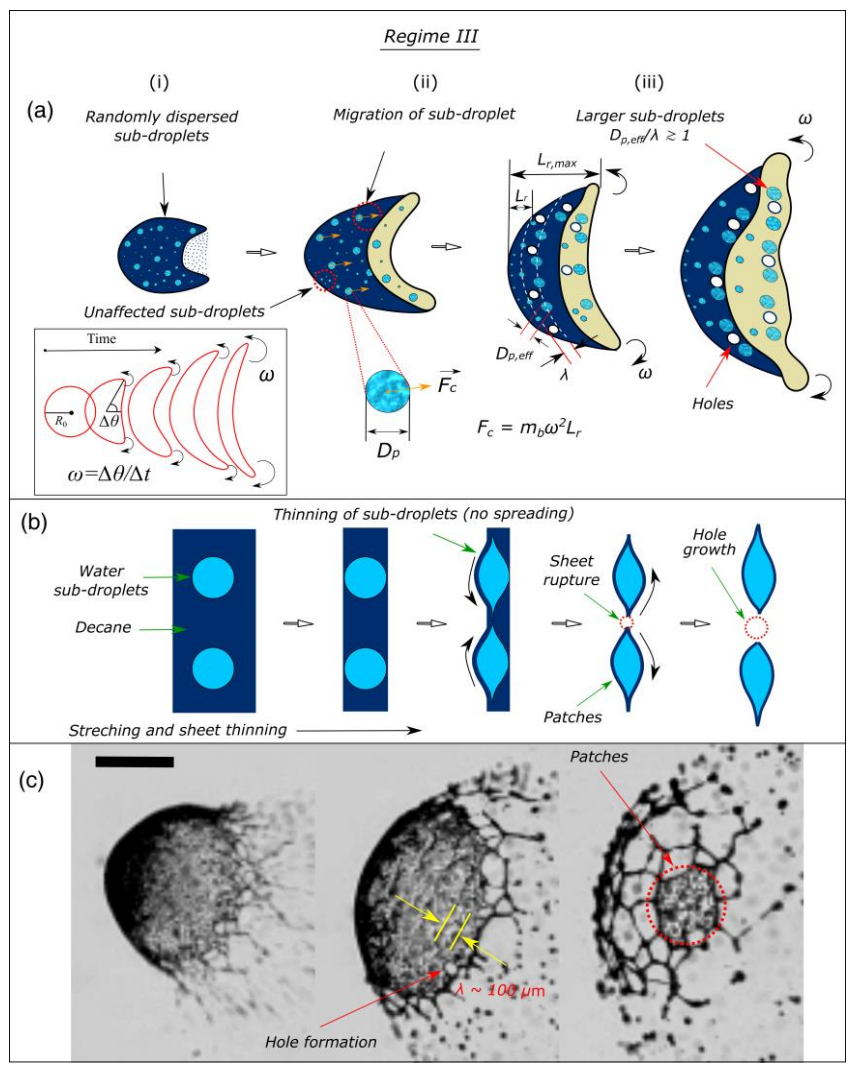

**Fig. 5. (a) Illustration of Regime III, indicating (i) randomly dispersed water sub-droplets in the expanding crown, (ii) migration of sub-droplets, and (iii) migrated larger sub-droplets and formation of holes surrounding them. Here $R_0$ is the radius of the droplet at the onset of breakup, $\theta$ is the angular displacement, and $\Delta t$ is the change in time. The inset illustrates the angular motion of the sheet edge with respect to the axis of symmetry. (b) Schematic describing the mechanism of sheet rupture and hole formation. (c) Expansion of liquid sheet and observation of wave-like pattern on the liquid sheet along with patches and holes.**

In the present case, the significant thinning of the liquid sheet implies that the water sub-droplets also undergo sufficient stretching. As a result, patch-like structures are visible on the sheet due to the surface tension gradient. The breakup mechanism observed in the present work is different from the sheet-thinning breakup reported by Minakov et al. [54]. Quantitatively, however, a valid comparison between the breakup mechanisms cannot be made since the sheet dynamics in the present study is induced by bubble breakup event. In contrast, sheet-thinning reported by Minakov et al. [54] is the result of aerodynamic force-induced breakup.

A probable hypothesis for the occurrence of the patches and holes (via sheet rupture) is explained through three stages, as shown in Fig. 5(a). Here, stage (i) corresponds to the presence of randomly dispersed water sub-droplets at the onset of the breakup of parent droplet. The size of the water sub-droplets for the emulsion mixture studied is observed to be rich in the range of 10–50 µm (~65%), followed by droplets in the range of 50–90 µm (~15%) and 90–130 µm (~15%) (See Fig. 6a).

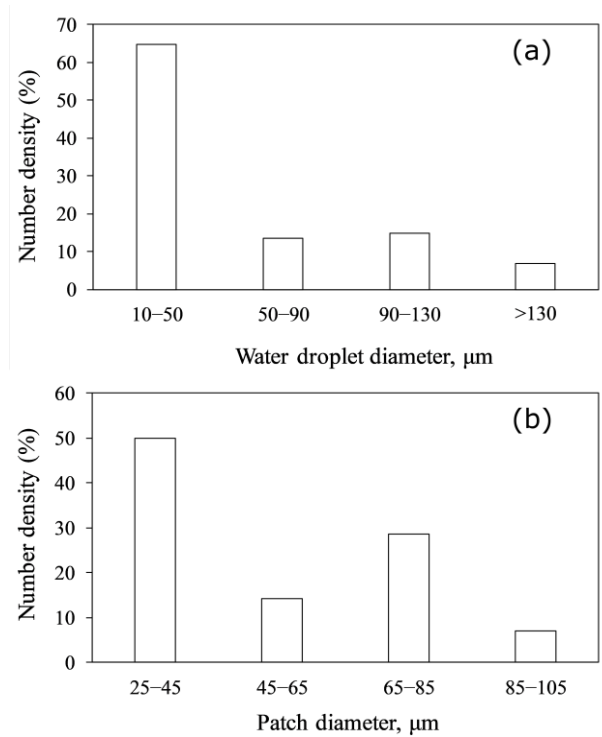

**Fig. 6. (a) Diameter distribution of water sub-droplets and (b) diameter distribution of patches.**

Stage (ii) represents the migration of these randomly dispersed sub-droplets towards the sheet edge due to sheet rotation (see the rotation in Fig. 4c). The water droplets undergo coalescence during migration in stage (ii).

The expansion and rotation of the liquid sheet is expected to induce centrifugal force which acts on the water sub-droplets as

$$F_c = m_{sd}\omega^2 L_r \qquad (2)$$

Here $m_{sd}$ represents the mass of sub-droplet, $\omega$ is the angular velocity of the expanding sheet, and $L_r$ is the radial length (see Fig. 5(a)). For a sub-droplet with a diameter of 25 µm, experiences a centrifugal force of ~ 1.3 x $10^{-7}$ N, while a sub-droplet with size 110 µm experiences a force of ~ 4 x $10^{-7}$ N. This implies that the water sub-droplets will be preferentially dispersed along the radial length of the liquid sheet. The bigger droplets will migrate towards the edge while the smaller ones will reside near the core. The water droplets near the sheet edge experience two orders of magnitude larger centrifugal force than the weight of the crown ($w_{crown} = m_{crown}g$) (Fig. 7). This spontaneous inhomogenization plays an important role in sheet breakup. As a consequence of this, small to large-sized droplets disperse proportionately along increasing radial length $L_r \propto F_c / w_{crown}$ (Fig. 7). The uncertainty in the determination of centrifugal force is ± 0.2 x $10^{-7}$ N, which arises primarily due to the error in the pixel measurement associated with sub-droplet size, radial length, and angular velocity.

Stage (iii) exhibits the appearance of dark patches on the liquid sheet that subsequently undergoes disintegration, forming holes (Fig. 7, left inset). Interestingly, the dimension of the dark patches seems to correlate with the order of water sub-droplets (Fig. 6(b)), indicating that the sub-droplets do not undergo much thinning. It is seen that the largest patches exist at the liquid sheet edge (large radial distances) while the smaller ones are discernible near the core. The size of these water sub-droplets can be correlated with the size of patches on the sheet, as seen in Fig. 6(b), suggesting that the water sub-droplets are, in fact, visible as dark patches. It is also probable that the presence of larger patches near sheet edge is the manifestation of higher outward acceleration and stretching of the sheet edge. However, further investigation is needed to validate this conjecture.

Figure 7 shows the linear growth of the normalized patch diameter ($D_p/\lambda$) with normalized centrifugal force ($F_c/w_{crown}$). Here, $\lambda$ is the radial distance between the consecutive rows of patches, which incidentally also matches the wavelength of the wave-like pattern reported in Fig. 5(c). Stage (iii), therefore, indicates the existence of sub-droplets as dark patches with the larger ones present near the sheet edge, whose surrounding regions rupture into circular holes.

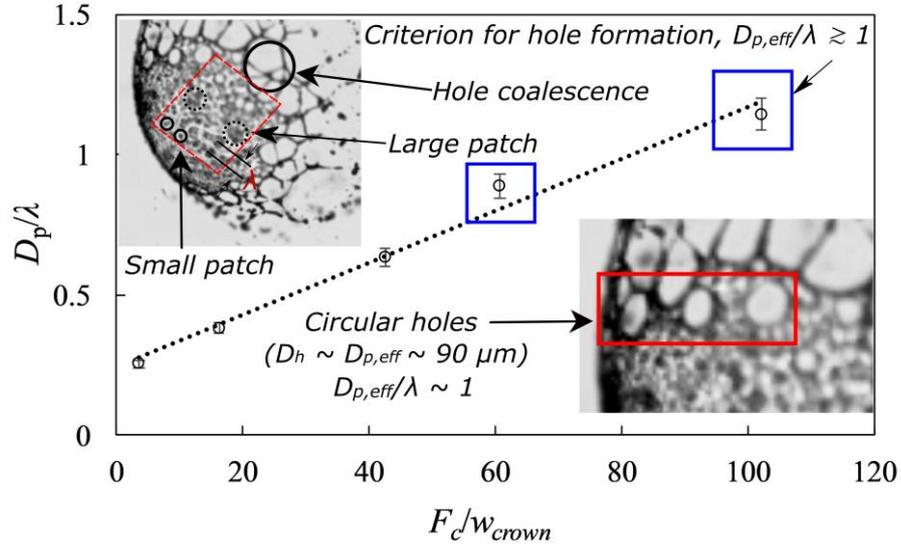

**Fig. 7. Variation of normalized patch diameter ($D_p/\lambda$) with the normalized centrifugal force ($F_c/w_{crown}$) along the radial distance of the sheet ($L_r$) at 100 μs from the onset of the breakup. The dotted line connecting the markers represents a linear trendline. The rectangular boxes on the curve signify the fulfillment of the criterion for hole formation ($D_{p,\,eff}/\lambda \gtrsim 1$). The inset on the left side displays small patches, large patches, and hole coalescence. The dotted rectangle indicates the region of interest for the evaluation of patch sizes. The inset on the right side highlights the circular holes of the size comparable to patch diameter near the sheet edge.**

A plausible mechanism associated with sheet rupture and hole formation is illustrated in Fig. 5(b). When the liquid sheet expands and thins, the region adjacent to the patches (representing water sub-droplets) is expected to rupture due to the lower surface tension of oil (decane) compared to that of water. It is known that, for an oil-in-water emulsion, the oil component spreads on water due to a positive spreading parameter [31,32]. In the present case, however, since water sub-droplets do not spread over decane (oil component), the thinning sheet encompassing the water sub-droplets undergo rupture. A similar hypothesis for sheet rupture and hole formation was also put forward by Aljedaani et al. [34] for a droplet with higher surface tension impacting on a thin liquid film of lower surface tension. In all our observations, it is observed that the rupture of the crown sheet through hole formation occurs when,

$$D_{p,\,eff}/\lambda \gtrsim 1 \qquad (3)$$

where $D_{p,\,eff}$ indicates effective patch diameter required for hole formation to occur on the sheet (see the illustration in Fig. 5(a)). Due to the curvature effects, the patches do not always remain in the same plane with respect to the camera, resulting in uncertainty in the measurement of patch diameter. However, the maximum uncertainty in the measurement of $D_p$ is within ± 5%. Figure 5(c) shows the experimental observation of the sheet expansion accompanied by the formation of patches and holes. A wave-like pattern is also observed on the sheet surface, whose wavelength is of the order of 100 µm. It should be noted that due to the complexity in capturing the focused sheet after the droplet breakup, clear segregation of patches, holes, and wave-like pattern could not be observed simultaneously in all the test cases.

The left inset of Fig. 7 shows multiple holes near the edge of the sheet encompassing web of complex ligament structures. Despite the stochastic behavior of sheet breakup, multiple evidences of hole formation during sheet expansion are obtained.

The holes exhibit rapid growth after its inception. The growth rate is observed to be in the range of 1.2 – 1.4 m/s. The growth rate can be further predicted using Taylor-Culick law [55,56] assuming uniform and constant sheet thickness as,

$$V_H = \sqrt{2\sigma/\rho h}. \tag{4}$$

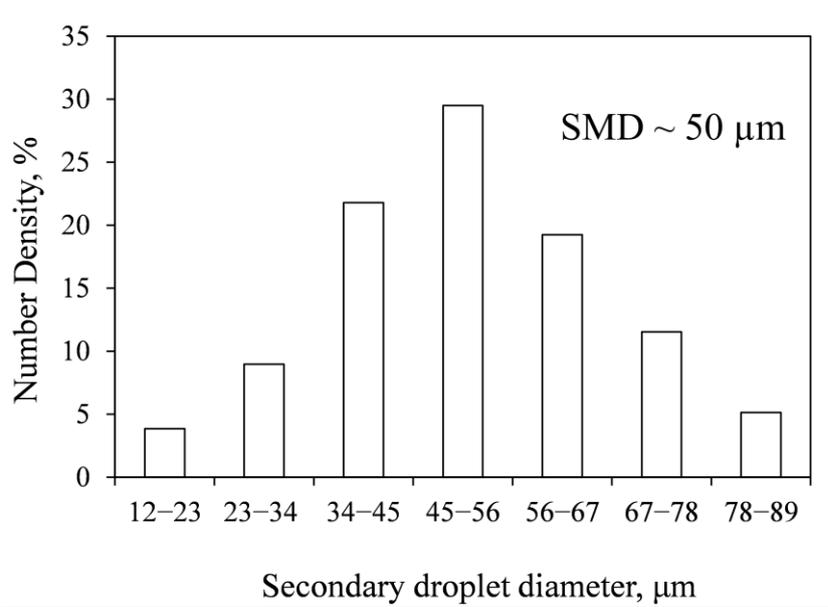

Fig. 8. Diameter distribution of secondary droplets originating from the disintegration of the crown sheet.

The order of magnitude of the theoretical hole growth rate ($V_H$ = 1.5 m/s) is comparable to the experimentally observed values. Similar values of hole growth rate (1.5 – 2.3 m/s) have also been reported in the literature [31,33] in the context of drop impacts. The ligaments originating from the hole formation are approximately uniform in diameter and undergo breakup into secondary droplets. Figure 8. Shows the size distribution of secondary droplets originated from the breakup of droplets corresponding to four test cases. The size distribution of the secondary droplets (which is ~ 50 µm) is also expressed in terms of Sauter mean diameter ($D_{32}$):

$$D_{32} = \frac{\sum n_i D_i^3}{\sum n_i D_i^2} \qquad (5)$$

Here, $D_i$ is the secondary droplet diameter, and $n_i$ is the number of secondary droplets possessing the same $D_i$ value. SMD is frequently used to express the drop size distribution of the secondary droplets due to the breakup of single droplets [57].

The initial development of the hole near the sheet edge is almost circular (right inset of Fig. 7); however, once the holes grow or coalesce, polygonal structures are formed with long filaments. The secondary droplets move at a significant forward velocity (~4 – 8 m/s) after their dissociation from the ligaments. The ligaments (aspect ratio ≥ 3.5) formed from the sheet become unstable and undergo Plateau-Rayleigh instability with a short capillary timescale ($\tau_c = 20$ µs).

# 4  Conclusions

We have provided a phenomenological description of breakup dynamics of an evaporating levitated emulsion droplet. The prominent conclusions derived from the present work are as follows:

1. The droplet life history consists of a) preheating and initial regression (~0.25 s), b) bubble formation and collapse ($10^{-5}$ s), and c) expansion of the liquid sheet with the formation of patches and subsequent rupture of thin sheet surrounding the patches ($10^{-4}$ s).
2. The collapse of the high-pressure vapor bubble results in the breakup of the droplet leading to the acceleration (~ $6 \pm 2 \times 10^4$ m/s$^2$) and expansion of a crown-like sheet. The area of this expanding crown varies linearly with the square of the time.
3. It has been hypothesized that the centrifugal force (~$10^{-7}$ N) driven migration naturally segregates the sub-droplet distribution leading to larger patches at the sheet edge.
4. Due to significant stretching and thinning, the localized regions of the sheet adjacent to these patches results in the creation of holes in the crown sheet. The holes are observed to exhibit rapid growth, and the growth rate is observed to be in the range of 1.2 – 1.4 m/s.
5. The ligaments generated from hole formation are observed to be uniform in diameter and undergo breakup into secondary droplets with SMD ~ 50 µm.